\documentclass[sigconf]{acmart}

\usepackage{wrapfig}
\usepackage{makecell}
\usepackage{verbatim}

\copyrightyear{2025} 
\acmYear{2025} 
\setcopyright{cc} 
\setcctype{by} 
\acmConference[UMAP '25]{33rd ACM Conference on User Modeling, Adaptation and Personalization}{June 16--19, 2025}{New York City, NY, USA} 
\acmBooktitle{33rd ACM Conference on User Modeling, Adaptation and Personalization (UMAP '25), June 16--19, 2025, New York City, NY, USA} 
\acmDOI{10.1145/3699682.3728318} 
\acmISBN{979-8-4007-1313-2/2025/06}

\begin{document}

\title{Generative Framework for Personalized Persuasion: Inferring Causal, Counterfactual, and Latent Knowledge} 

\author{Donghuo Zeng}
\authornote{Both authors contributed equally to this research.}
\affiliation{%
  \institution{KDDI Research, Inc.}
  \city{Saitama}
  \country{Japan}
}
\email{do-zeng@kddi-research.jp}

\author{Roberto Legaspi}
\authornotemark[1]
\affiliation{%
  \institution{KDDI Research, Inc.}
  \city{Saitama}
  \country{Japan}
}
\email{ro-legaspi@kddi-research.jp} 

\author{Yuewen Sun}
\affiliation{%
  \institution{MBZUAI \& CMU}
  \city{Abu Dhabi \& Pittsburgh}
  \country{ }
}
\email{yuewen2819@gmail.com}

\author{Xinshuai Dong}
\affiliation{%
  \institution{Carnegie Mellon University}
  \city{Pittsburgh}
  \state{Pennsylvania}
  \country{USA}
}
\email{xinshuad@andrew.cmu.edu}

\author{Kazushi Ikeda}
\affiliation{%
  \institution{KDDI Research, Inc.}
  \city{Saitama}
  \country{Japan}
}
\email{kz-ikeda@kddi-research.jp}

\author{Peter Spirtes}
\affiliation{%
  \institution{Carnegie Mellon University}
  \city{Pittsburgh}
  \state{Pennsylvania}
  \country{USA}
}
\email{ps7z@andrew.cmu.edu}

\author{Kun Zhang}
\affiliation{%
  \institution{MBZUAI \& CMU}
  \city{Abu Dhabi \& Pittsburgh}
  \country{ }
}
\email{kunz1@andrew.cmu.edu}

\renewcommand{\shortauthors}{Zeng, Legaspi, Sun, Dong, Ikeda, Spirtes \& Zhang}

\begin{abstract}

We hypothesize that optimal system responses emerge from adaptive strategies grounded in causal and counterfactual knowledge.  
Counterfactual inference allows us to create hypothetical scenarios to examine the effects of alternative system responses. 
We enhance this process through causal discovery, which identifies the strategies informed by the underlying causal structure that govern system behaviors.
Moreover, we consider the psychological constructs and unobservable noises that might be influencing user-system interactions as latent factors. We show that these factors can be effectively estimated. 
We employ causal discovery to identify strategy-level causal relationships among user and system utterances, guiding the generation of personalized counterfactual dialogues. We model the user utterance strategies as causal factors, enabling system strategies to be treated as counterfactual actions. Furthermore, we optimize policies for selecting system responses based on counterfactual data. Our results using a real-world dataset on social good demonstrate significant improvements in persuasive system outcomes, with increased cumulative rewards validating the efficacy of causal discovery in guiding personalized counterfactual inference and optimizing dialogue policies for a persuasive dialogue system.
\end{abstract}




\maketitle

\section{Introduction} \label{Intro} 

Persuasive technologies (PTs) are created to bolster, modify, or fashion user opinions, attitudes, and behaviors in a non-coercive and honest manner~\cite{Fogg2003, OinasKukkonen2009}. Many studies have shown that PTs are effective for various purposes, inter alia, health intervention~\cite{Kelders2012, Orji2018, Asbjørnsen2019, Furumai2024}, marketing~\cite{Braca2023}, energy usage and conservation~\cite{Chiu2020, Böckle2020, Wu2020}, and encouraging prosocial behavior~\cite{Wang2019, Oliveira2021, Furumai2024}. Yet, PTs often struggle by not fully knowing at design time the varying user behaviors and contexts concerned~\cite{Kaptein2015}, further exacerbated by the dearth of observed training data for the persuasion task. Furthermore, traditional persuasive systems often rely on predefined strategies or static models that fail to account for the dynamic and complex nature of user behavior, leading to suboptimal persuasion outcomes.

\begin{figure}[ht]
	\centering
		\includegraphics[width=0.93\linewidth]{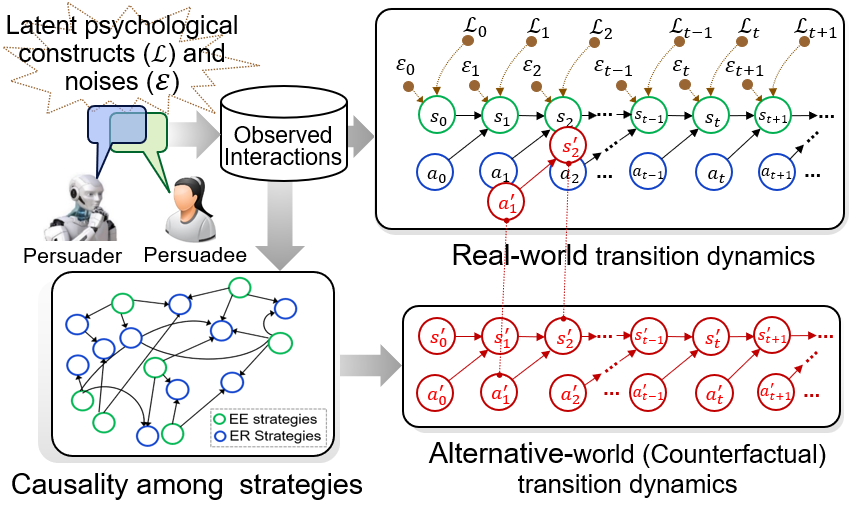}
        \vspace*{-2mm} 
	  \caption{
      \textbf{Schema of our hypothesis}. 
      An observed state transition $\{(s_{t}, a_{t}, s_{t+1})\}^{T-1}_{t=0}$, 
      where $s_t$ is the user state at time $t$, $a_t$ is the action taken by the system, and $s_{t+1}$ is the next user state, follows the rules of a structural causal model~\cite{Pearl2009} while accounting for hidden psychological constructs ($\mathcal{L}$) and unobserved noises ($\varepsilon$). Identified causal relations among strategies
      make it easier to draw well-founded counterfactual states and actions
      to achieve the desired persuasion outcomes.
      }
      \vspace{-6mm}
    \label{fig:theoretical_schema}
\end{figure}

We hypothesize that optimal system actions would emerge as by-products of adaptive strategies anchored in causal and counterfactual knowledge. This knowledge captures the underlying causal structures, hidden psychological constructs, and unobserved noises that influence user-system state transition dynamics (Fig.~\ref{fig:theoretical_schema}).
Unlike previous methods that strictly hang on to predefined, mostly theory-based, persuasive strategies, we identify from observational data the causal strategies governing user behaviors, as well as the elicited system behaviors as causal effects.
We leverage counterfactual inference~\cite{hoch1985counterfactual,Pearl2018,Bareinboim2022} to generate hypothetical scenarios to explore and understand what would happen if certain actions or conditions were different. 
It generates new reachable data points by manipulating input variables to analyze and explore potential outcomes, thereby not only augmenting the existing data but also enhancing data quality tailored to task objectives~\cite{Lu2020, Pitis2020}. 
Thus, counterfactual inference focuses on diversifying existing data while explicitly considering the transition dynamics of states, actions, and next states~\cite{Lu2020, Sun2024}. Our use of causal discovery ensures that data are augmented and diversified per edicts of the underlying causal relations, thereby facilitating a principled counterfactual inference. 


In addition, we account for latent factors - hidden psychological constructs ($\mathcal{L}$) and unobserved noises ($\varepsilon$) - possibly influencing user-system interactions.  Many constructs in psychology are often modeled as hidden variables in theories of human behavior, such as intelligence~\cite{Spearman1904}, spatial ability~\cite{Thorndike1921}, sense of agency~\cite{Legaspi2024}, and personality traits~\cite{Goldberg1990,Costa1992}.
However, it is also the case that during experimentation, unobservable variability, randomness, or idiosyncratic influences, occur as noises, which are not attributable to the latent psychological variables themselves.

To explore our hypothesis, we formulate an approach to realizing a persuasive dialogue system in which its persuasiveness largely depends on the ability to adapt its communication strategies to the causal latent factors and evolving states of its users.
Specifically, we employ the Greedy Relaxation of the Sparsest Permutation (GRaSP)~\cite{lam2022greedy} algorithm to uncover cause-effect relations at the utterance-strategy level.
We investigate two counterfactual inference methods while extending their original versions: Bi-directional Conditional Generative Adversarial Network (BiCoGAN)~\cite{jaiswal2019bidirectional,Lu2020}, and kernel quantile regression (KQR) for counterfactual inference~\cite{Xie2024}. Both models estimate the structural causal model, but while BiCoGAN explicitly estimates the noise terms, KQR bypasses the need of estimating the noises. Furthermore, we develop our Turn-based Persuadee Personality Prediction Model (TP3M) to estimate latent user personality traits to foster user-centered adaptive dialogues 
as the persuasive utterances are adjusted 
following the predicted traits. 
At the last stage of our pipeline, we employ the Dueling Double Deep Q-Network (D3QN)~\cite{raghu2017deep} model to optimize the selection of system utterances based on the counterfactual data. The system becomes better equipped to dynamically adapt its strategies by learning optimal interaction policies from the causal discovery-navigated counterfactual data construction, ultimately improving the effectiveness of its persuasive responses. 

For our experiments, we use a real-world dataset in which one individual (persuader) successfully influenced (or not) another individual (persuadee) to make a donation to charity. Our experimental results demonstrate the ability of our novel approach, in contrast to baseline methods that are grounded in the work we reported in~\cite{zeng2024counterfactual}, to significantly improve the cumulative rewards that represent the goodness of the persuasion outcome (i.e., amount of donation given). 
Our findings underscore the potential of integrating user-personalized and adaptive causal discovery, counterfactual inference, and latent factor estimation to enhance persuasive action policies and improve system outcomes. 

\section{Related Works} \label{related_works}
\begin{figure*}[t!]
    \centering
    \includegraphics[width=0.95\textwidth]{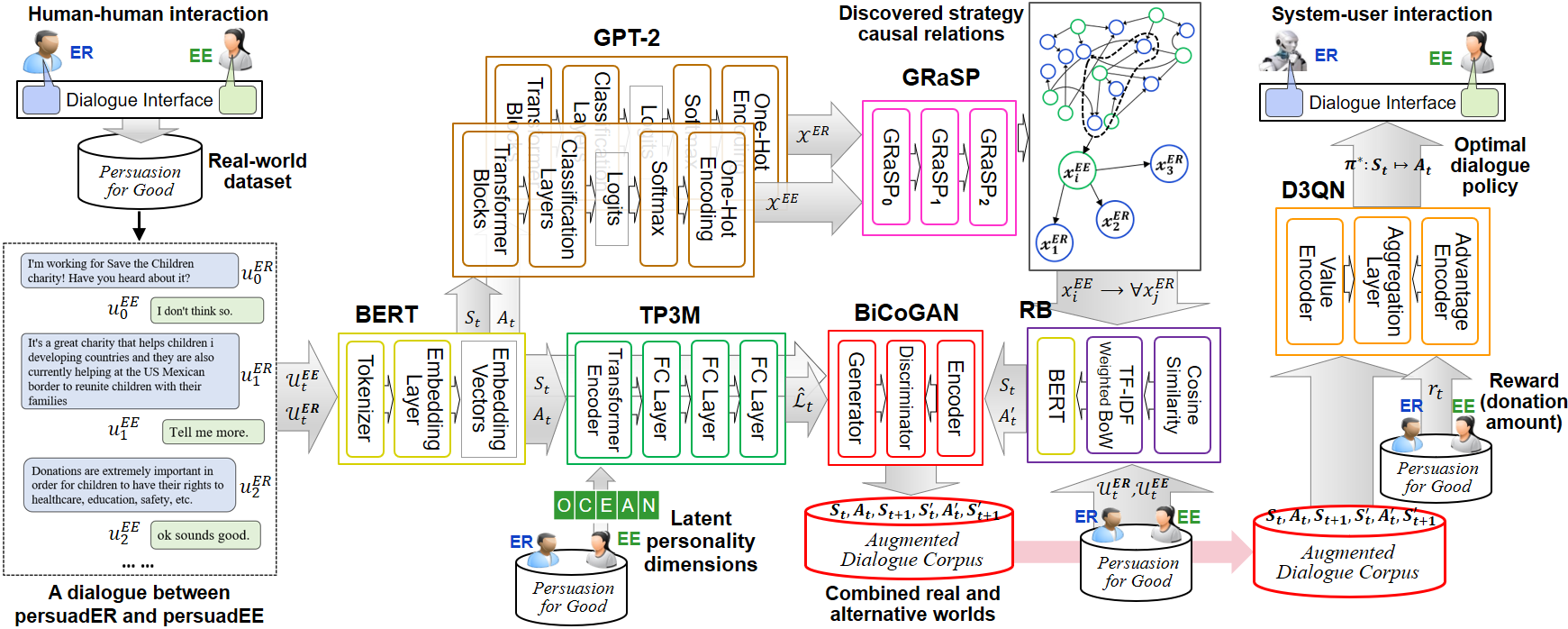}
    \caption{\textbf{Generative framework for personalized causality-driven counterfactual persuasions.} We represent the persuadEE and persuadER utterances as BERT-embedded state-action sequences. TP3M recovers the latent personality dimensions using the state-action sequences and the associated EE OCEAN values in every dialogue. The strategies associated with the utterances are obtained by fine-tuning two GPT-2 models, and the causal relationships between EE and ER strategies are discovered by GRaSP, which will then guide the construction of counterfactual actions. The counterfactual data are created using BiCoGAN, or via kernel quantile regression (not shown here but in Fig.~\ref{fig:kqr_model}), by utilizing the estimated latent personality dimensions and the counterfactual actions obtained by the RB model. Finally, D3QN learns the optimal system response policies, with the donation amounts as reward values, to improve persuasion outcomes.
    }
    \vspace{-3mm}
    \label{fig:system_architecture}
\end{figure*}

Most of science revolves around identifying causal structures and the laws or patterns that govern them~\cite{Glymour2019}. 
Causal discovery, which aims to uncover hidden causal structures from observed data, has been used for decades~\cite{EbertUphoff2014} and has proven effective in a wide array of fields~\cite{spirtes2000causation, Glymour2019, Pearl2019, Runge2019, Assaad2022}. 
It is the case, however, that not all causal outcomes can be observed; hence, cause-effect relations may not be simply computed even in the best scenario (unless, say, we go back in time). 
Nonetheless, experimental and simulation methods for estimating causality demand conceptual clarity about alternative causal conditions
by comparing outcomes resulting from being exposed to alternative causal states \textit{ceteris paribus}. 

Counterfactual reasoning~\cite{hoch1985counterfactual,Pearl2009} is the conceptual or cognitive process of leveraging hypothetical scenarios that are alternatives to reality. It poses \textit{what if} questions to understand the alternative cases based on changes to past events. This is applied formally as counterfactual inference~\cite{Pearl2009, Pearl2018,Pearl2019,Bareinboim2022} to computationally generate the alternative scenarios that could have led to different outcomes.
This can provide revealing insights into causal discovery results~\cite{spirtes2000causation, Pearl2009, Pearl2018, Pearl2019}. 
However, conventional approaches to counterfactual inference often rely on having access to or estimating a structural causal model (SCM)~\cite{Bareinboim2022}. Unfortunately, the SCM is often unavailable~\cite{Xie2024}. Moreover, current methods often rely on specific distributional assumptions regarding the noise terms, thereby restricting the model class. 
To tackle these issues, various approaches have been proposed for estimating the SCM using observational data~\cite{Lu2020, Xie2024, Sun2024}. For instance, the 
BiCoGAN~\cite{Lu2020} learns the SCM by minimizing via a discriminator the disparity between the input real data and the generated data, maintaining realistic counterfactual states that are aligned with the observed scenarios. Simultaneously, it also estimates the value of the noise term that represents the disturbances arising from unobserved factors. On the other hand, the work in~\cite{Xie2024} reframes counterfactual inference as an extended quantile regression problem, evading the estimation of the noise conditional distributions.

When focused on PTs, our work stands out even more, as we address the generation of personalized and causally-informed counterfactual strategies. Personalization is important for any PT to adapt to user-specific contexts to enhance system effectiveness~\cite{kaptein2015personalizing,hirsh2012personalized,orji2018personalizing,rieger2022towards,matz2023potential}, for instance, through persuasion profiling~\cite{kaptein2015personalizing}, or with a dynamic user model based on dialogue history and the user's inclination to be persuaded by the system~\cite{tran2022ask}.
We, on the other hand, focus on leveraging latent personality traits~\cite{Goldberg1990,Goldberg1992,Costa1992,Costa2017,SunHYZDJSLISZ24}. 
We also address the inability to directly handle noises (e.g., conflicting or ambiguous user feedback and environmental disruptions) that introduce randomness and variability in system behavior. 
Indeed, a major limitation of current PT approaches is the inadequate modeling of latent psychological constructs and other unobserved personal factors influencing user responses~\cite{Borsboom2003, Ndulue2022}. 
Furthermore, most PTs derive their strategies from theoretical frameworks~\cite{Fogg2003, Cialdini2006, OinasKukkonen2009} that offer templates for influencing user behavior, or rely on user interaction data to derive interaction rules~\cite{Riley2011,NahumShani2018,Shin2018,Dalecke2020}. However, systems that lack flexibility, as they remain reliant on predefined strategies and the limited amount of observed data, fall short of effectively adapting to the evolving, nuanced, and dynamic nature of user behavior. The reliance on static models, predefined strategies, or insufficient data often overlooks the causal and counterfactual dynamics governing real-world user-system interactions~\cite{Pearl2009, Xie2024,zeng2025causal}.

Against the above computational and application backdrop, we forge our framework (illustrated in Fig.~\ref{fig:system_architecture}) that generates personalized, counterfactual data to achieve generalizability, address data scarcity, and generate system responses that could have occurred in an alternative reality~\cite{zeng2024counterfactual}. Our choice of the methods comprising our generative framework is driven by their sufficiency in testing and validating our hypothesis~\footnote{We detail in ~\textbf{Supplementary Materials (\url{https://github.com/ZenzenDatabase/PersonPersuasion})} our motivations and justifications for choosing these methods.}.
To our best knowledge, no persuasive dialogue system, nor PT in general, has used this kind of approach to improve their persuasive interaction outcomes.

\section{Generative Framework for Personalized Causality-driven Counterfactual Persuasions} 
\label{syst_architect}

\subsection{Representing Dialogue Utterances as State-Action Sequences}
Our framework begins with utilizing BERT embeddings, which are extracted by a pre-trained BERT model~\cite{Devlin2019} (Fig.~\ref{fig:system_architecture}, in yellow), to represent the dialogue utterances in the \textit{Persuasion for Good} (referred to here as P4G) dataset~\footnote{This dataset can be readily accessed in \textbf{\url{https://convokit.cornell.edu/documentation/persuasionforgood.html}}, in which 300 dialogues contain human annotations.}~\cite{Wang2019}.
P4G is a collection of 1017 human-human persuasion dialogues.
During the actual online tasks, one participant was tasked as persuader (ER) to convince the other who acted as persuadee (EE) to make an actual donation to the charitable organization called \textit{Save the Children}~\cite{saveforchildren}.
We employ the trained BERT model to convert every dialogue utterance made by EE and ER, $u^{EE}_t$ and $u^{ER}_t$, to their 768-dim BERT embeddings, and assign these embedded vectors, respectively, as $s_t$ and $a_t$ of the state-action sequences. 
We treat the P4G dataset as ground truth (real world) data, and then use our framework to augment it with counterfactual (alternative world) data. The end in mind is to capitalize on the learned optimal persuasion dialogue policy ($\pi^*$) when interacting with users at deployment so as to maximize the persuasive system outcomes (Fig.~\ref{fig:system_architecture}, top right end). It is not within the scope of this paper, however, to realize the actual dialogue interface that facilitates the system-user interactions. Rather, to equip such an interface with the intelligence that would emerge from our generative framework. 

\subsection{Recovering the Latent Personality Traits}
Known for its five broad trait dimensions - 
Openness, Conscientiousness, Extraversion, Agreeableness, and Neuroticism (or OCEAN) - the Big Five is a well-established classification system for personality traits~\cite{Costa2017, Wright2017, DeYoung2007, Xu2022}. 
Variations in this small set of broad dispositional traits linked to social life often predict significant differences in behavior, especially when behavior is considered across different situations~\cite {mcadams2006new}. For our computational purposes, we shall use OCEAN to refer to these five latent personality traits.    

Estimating the OCEAN values
helps in dynamically adjusting the persuasive system's utterances based on the inferred user personality traits to achieve a user-centered dialogue. For this purpose, we developed TP3M (Fig.~\ref{fig:system_architecture}, in green), which is comprised of a Transformer encoder~\cite{Vaswani2017Transformer} followed by three fully connected (FC) layers. 
We trained TP3M using the BERT-embedded utterances, alongside the annotated OCEAN values of EE (included in P4G), as ground truth
~\cite{Wang2019}. 
However, since counterfactual dialogues would be generated, we need TP3M to recover from the generated alternative-world utterances the OCEAN values, as follows:
\begin{equation}
\vspace{-1mm}
    \mathcal{\hat{L}}_t = tp3m(A'_{t-1}, S'_{t}),
    \label{eq:tp3m_func}
    \vspace{-0.5mm}
\end{equation}
where $\mathcal{\hat{L}}_t$, is a five-dimensional vector that indicates the values of OCEAN (e.g., O: 3.8, C: 4.4, E: 3.8, A: 4, N: 2.2).
In other words, we posit that EE's response is influenced by how EE's personality perceives ER's response. For example, an introverted EE may respond negatively to ER saying, ``You can meet our other volunteers!''
Hence, during its persuasive dialogues with the user, the system would estimate the OCEAN values to better understand the user state $s_t$ during the dialogue. This enables the system to dynamically adapt its counterfactual utterance $a'_t$ based on $s'_t$ and $\mathcal{\hat{L}}_t$.

\subsection{Mapping Embedded Utterances to Strategies}
\label{strategy_mapping}
We employ two fine-tuned GPT-2 models to understand the strategies behind the utterances (Fig.~\ref{fig:system_architecture}, in brown). 
Processing the BERT-embedded utterances using the Transformer and classification layers of a fine-tuned GPT-2~\cite{radford2019language} produces a vector of logits. Let variable $\mathbf{l}$ represent the logits for the set of strategies, i.e., $\mathbf{l} = [l_1, l_2, l_3, ..., l_i, ..., l_K]$, where $l_i$ is the logit score for strategy $i$ and $K$ is the number of strategies ($K^{EE}=23$, $K^{ER}=27$). We convert the logits into probabilities using softmax. Specifically, the softmax function for logit vector $\mathbf{l}$ is defined as $\textit{softmax}(\mathbf{l})_i = e^{l_i}/{\sum_{j=1}^{K} e^{l_j}}$.
We then use one-hot encoding to specify the most probable strategies that correspond to the utterance. In other words, we fine-tune two separate GPT-2 models for the two roles and thus have the two fine-tuned GPT-2 as final models, which we then use to generate the softmax probabilities to identify the most probable strategies. This process therefore maps $(s_t, a_t)$ to $(x^{EE}_t$, $x^{ER}_t)$, in which $x^{EE}_t$ and $x^{ER}_t$ are strategies.

Out of the 1017 dialogues in P4G, a subset of 300 dialogues had been annotated (referred to as ANNSET), while the remaining 717 dialogues were left unannotated. The utterances in ANNSET were annotated with 50 distinct strategies under four categories: \textit{appeal} (7), \textit{inquiry} (3), \textit{task-related non-persuasive} (17) strategies for ER, and \textit{general conversation behavior} (23)  strategies for EE.
We train the GPT-2 models using ANNSET to identify the strategies that underlie the utterances. We then use the trained models to predict the strategies of the 717 unannotated dialogues. We validated the models' effectiveness using five-fold cross-validation with ANNSET, achieving an accuracy of 92.3\%. We also measured the intra- and inter-strategy similarities~\cite{soler2013data} between utterance pairs within the same strategy (intra-), and among centroids of the utterances from different strategies (inter-) in both the ANNSET and the unannotated dialogues whose strategies had been predicted. More details about the annotation process can be found in~\cite{Wang2019}.
We can see from Table~\ref{tab:similarities} that under the same metric, the values for both sets are close to each other.
What these results demonstrate is the capability of the trained GPT-2 models to strongly predict the strategies underlying the utterances.

\begin{table}[h]
\small
\vspace{-2.5mm}
  \caption{Intra- and inter-strategy cosine similarities for both ANNSET and the unannotated dialogues.}
  \label{tab:similarities}
  \centering
  \vspace{-3mm}
  \resizebox{0.95\linewidth}{!}{
    \begin{tabular}{ccc} 
        \hline 
        Dataset & Intra & Inter \\ 
        \hline 
        ANNSET, with the annotated strategies & 0.632 & 0.559 \\ 
        Unannotated, using the predicted strategies & 0.645 & 0.576 \\ 
        \hline
    \end{tabular}
  }
\vspace{-3mm}
\end{table}

\subsection{Discovering Causality Among Strategies} \label{causal_discovery}
To ensure that the counterfactual utterances are generated in a principled manner, our framework leverages the discovered causal relations among strategies to help shape the formulation of counterfactual data. This ensures that the counterfactual actions are not from a vacuum nor selected at random, but are grounded in the discovered causal relations.
We hypothesize that causal relations exist between the EE and ER strategies, which influence the dialogue outcome. For each utterance in P4G, only one role (EE or ER) is active and the other has no associated strategy. Furthermore, each dialogue contains the donations of both EE and ER. However, since we envision the system to be the ER, and the user as EE, our primary focus in this work is the actual amount of EE's donation as it directly pertains to the persuasion outcome. A low to no donation can be indicative of ER's less effective persuasive strategies. 
By leveraging their causal structure, the strategies can be adjusted to maximize the final donation. 

We employ GRaSP~\cite{lam2022greedy} (Fig.~\ref{fig:system_architecture}, in pink) to identify the causal relations among the 50 strategies. GRaSP’s scalability and reliability make it valuable for this high-dimensional and sparse setting.
Given its three-tiered relaxation framework, moving from a lower to a higher tier results in a gradual but effective theoretical relaxation of the permutation search space, yielding improved accuracy. GRaSP has proved particularly effective for dense causal graphs and large-scale models, outperforming many existing simulation algorithms~\cite{chickering2002optimal, ramsey2017million}. This makes GRaSP well-suited for inputs with many variables, such as the 23-dim $\mathcal{X}^{EE}$ and 27-dim $\mathcal{X}^{ER}$ strategy variables, and makes GRaSP a valuable tool for analyzing the causality among strategies. 

\subsection{Converting Strategies Back to Utterances } 
\label{retrieval_model}
With the discovered causal graph, $\mathcal{X}^{EE}$ (the causes) would serve to identify the corresponding $\mathcal{X}^{ER}$ (the effects). Specifically, given persuadee state $s_t$, which has been derived from persuadee utterance $u_t^{EE}$ with identified strategy $x_i^{EE}$, we ask which persuader strategy is best suited to respond to $x_i^{EE}$. Let us denote this relation as $x_i^{EE} \rightarrow \tilde{x}^{ER}$, with $\tilde{x}^{ER}$ being the most suitable response strategy of EE. The next question is how to generate the persuader utterance $u_t^{ER}$, which is in natural language, that best instantiates $\tilde{x}^{ER}$.
Given the cause-effect relations $x_i^{EE} \rightarrow x^{ER}_{j=1,2, ..n}$ in the discovered causal graph, where $n$ is the number of effect strategies influenced by the common cause strategy $x_i^{EE}$, we randomly select each time one cause-effect relation $x_i^{EE} \rightarrow x_j^{ER}$. We then use cosine similarity to score and select the most suitable effect strategy 
that would now correspond to $\tilde{x}^{ER}$. To convert $\tilde{x}^{ER}$ to an utterance, we rank all candidate persuader utterances $\{u^{ER}\} \subseteq \mathcal{U}^{ER}$ that could be associated with $\tilde{x}^{ER}$. This ranking is based on the measured similarity between each $u^{ER}$ that follows any $u^{EE}$ whose identified cause strategy is $x_i^{EE}$. We then randomly choose one of three $u^{ER}$s that got the highest similarity scores to become the counterfactual action $a^{'}_{t}$ in response to $s_t$. This random selection is to evade repetitively and greedily selecting the same utterance at the very top.

We implement the above process in a retrieval-based (RB) model (Fig.~\ref{fig:system_architecture}, in purple) that was validated in~\cite{tran2022ask}. Specifically, we use this model to identify the most suitable equivalent utterance $\mathcal{U}^{ER}_t$ (in natural language) for a given one-hot encoded strategy $\mathcal{X}^{ER}$, in response to $\mathcal{U}^{EE}_t$ with strategy $\mathcal{X}^{EE}$. 
The RB converts strategies to utterances by utilizing the TF-IDF weighted bag-of-words vector for the utterance representation. Finally, the RB model passes the BERT embedding of the resulting utterances $(\mathcal{U}^{EE}_t, \mathcal{U}^{ER}_t) \equiv (S_t, A'_t)$, which now include the selected counterfactual actions, on to the counterfactual inference methods.

\subsection{Counterfactual Inference} 
\label{counter_inf}

We perform counterfactual inference (CI) to explore the outcomes that might have arisen under different conditions. CI involves imagining alternative scenarios that differ from what was actually observed~\cite{Pearl2009, Pearl2018,Pearl2019,Bareinboim2022}. From the perspective of state transition dynamics, CI explains how the subsequent state would be affected had the system chosen an action different from the observed data (Fig.~\ref{fig:theoretical_schema}). We investigate two CI mechanisms: BiCoGAN (Fig.~\ref{fig:system_architecture}, in red) and CI-KQR (Fig.~\ref{fig:kqr_model}), being guided by causal relations at the strategy level and then explicitly estimating the noise terms or not.

\begin{figure}[h]
\vspace{-1mm}
    \centering
    \includegraphics[width=0.95\linewidth]{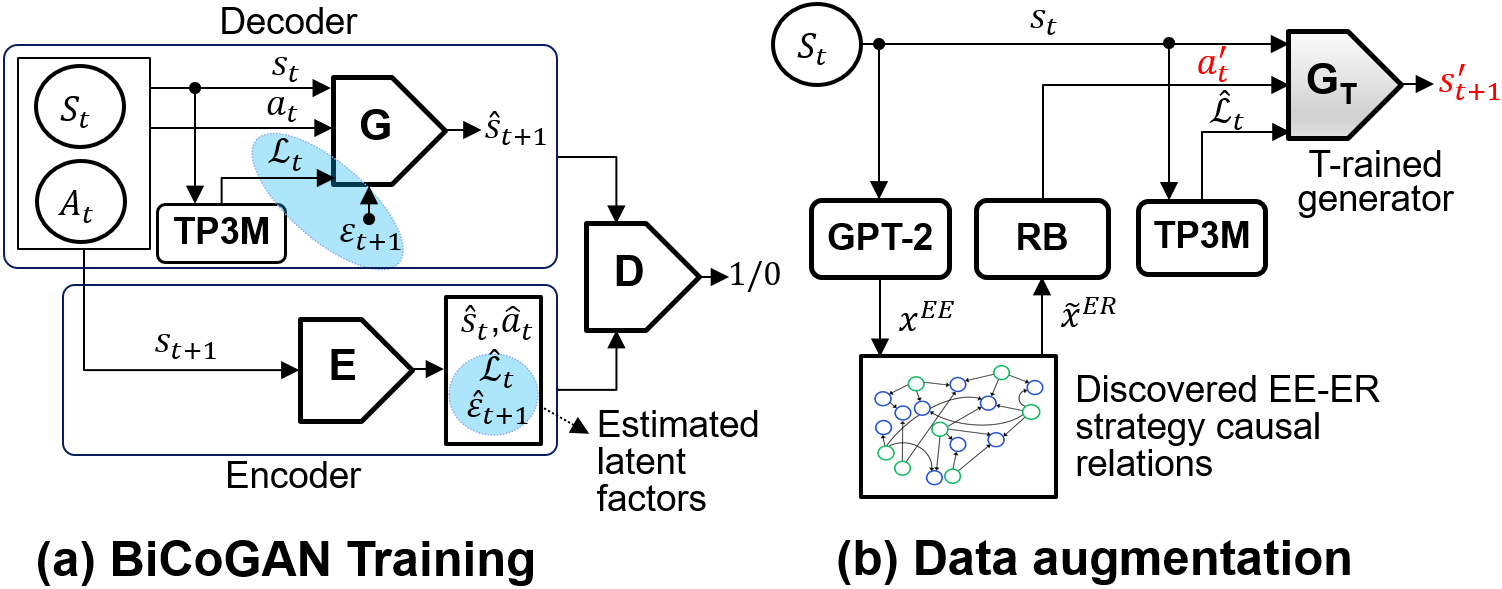}
    \caption{Novel extension of BiCoGAN. (a) Training with TP3M, generator $G$, encoder $E$, and discriminator $D$. (b) Discovered casual graph links the GPT-2 and retrieval-based models to generate via the trained generator $G_{T}$ the counterfactual action $a^{'}_{t}$ and next state $s^{'}_{t+1}$.
    } 
    \vspace{-5mm}
    \label{fig:causal_bicogan}
\end{figure}

\subsubsection{Novel extension of the Bidirectional Conditional GAN}
During training, in Fig.~\ref{fig:causal_bicogan}(a), BiCoGAN minimizes the disparity between real data and generated counterfactual data, ensuring that realistic counterfactual data align with the observed scenarios. It also estimates the noise term $\varepsilon_{t+1}$, thereby abstracting in it the influence of unobserved noise factors. We extend BiCoGAN by incorporating estimates of the latent personality dimensions into $\hat{\mathcal{L}}_{t}$.
Although $\hat{\mathcal{L}}_{t}$ is an estimate (cf: Eq.~\ref{eq:tp3m_func}), generator $G$ treats it as ground truth, which is justified by the high accuracy of $\hat{\mathcal{L}}_{t}$ (later in Section~\ref{results}). Thus, we denote the OCEAN values entering $G$ as 
$\mathcal{L}_{t}$ $\equiv$ $\hat{\mathcal{L}}_{t}$.


This extended BiCoGAN consists of two parts: 1) a generative model that maps ($s_{t}$, $a_{t}$, $\mathcal{L}_{t}$, $\varepsilon_{t+1}$) to $s_{t+1}$ via the generator $G$, and 2) an inference mapping from $s_{t+1}$ to ($\hat{s}_{t}$, $\hat{a}_{t}$, $\hat{\mathcal{L}}_{t}$, $\hat{\varepsilon}_{t+1}$). The discriminator $D$ is trained to discriminate between joint samples from the decoder and encoder distributions, respectively:
{\small
\begin{equation}
\begin{aligned}
    P(\hat{s}_{t+1}, s_{t}, a_{t}, \mathcal{L}_{t}, \varepsilon_{t+1}) &= P(s_{t}, a_{t}, \mathcal{L}_{t}, \varepsilon_{t+1})P(\hat{s}_{t+1}|s_{t}, a_{t}, \mathcal{L}_{t}, \varepsilon_{t+1}), \\
    P(s_{t+1}, \hat{s}_{t}, \hat{a}_{t}, \hat{\mathcal{L}}_{t}, \hat{\varepsilon}_{t+1}) &= P(s_{t+1})P(\hat{s}_{t}, \hat{a}_{t}, \hat{\mathcal{L}}_{t}, \hat{\varepsilon}_{t+1}|s_{t+1}),
\end{aligned}
\label{eq:causal_func}
\end{equation}
}
where $\hat{s}_{t}$, $\hat{s}_{t+1}$, $\hat{a}_{t}$, $\hat{\mathcal{L}}_{t}$, and $\hat{\varepsilon}_{t+1}$ are the estimates of $s_{t}$, $s_{t+1}$, $a_{t}$, $\mathcal{L}_{t}$, and $\varepsilon_{t+1}$.
To deceive the discriminator model, the objective function is optimized as a minimax game, defined as
{\small
\setlength{\abovedisplayskip}{5pt}  
\setlength{\belowdisplayskip}{5pt} 
\begin{equation}
\begin{aligned}
\min_G \max_D &V(D, G, E)= \min_G \max_D \{\mathbb{E}_{s_{t+1} \sim p_{\text{data}}(s_{t+1})}[\log D(E(s_{t+1}), s_{t+1})] \\ 
&+ \mathbb{E}_{z_{t} \sim p(z_{t})}[\log(1 - D(G(z_{t}), z_{t}))] \\
&+\lambda E_{(s_{t}, a_{t}, s_{t+1}) \sim p_{\text{data}}(s_{t}, a_{t}, s_{t+1})} [R((s_{t}, a_{t}), E(s_{t+1}))]\},
\end{aligned}
\end{equation}
}
where $z_{t} = (s_{t}, a_{t}, \mathcal{L}_{t}, \varepsilon_{t+1})$, and $R$ is regularized by the hyper-parameter $\lambda$ to prevent over-fitting.


We further extend BiCoGAN by constructing the counterfactual actions following the discovered causal graph, shown in Fig.~\ref{fig:causal_bicogan}(b). We predict strategy $\mathcal{X}^{EE}$ using the fine-tuned GPT-2 model and traverse the causal graph to select strategy $\mathcal{\tilde{X}}^{ER}$ in order to obtain the counterfactual action $a^{'}_t$ that has the highest contextual similarity 
(detailed in Sections~\ref{causal_discovery} and \ref{retrieval_model}). 
Finally, the tuple ($s_t$, $a^{'}_t$, $\hat{\mathcal{L}}_{t}$) is fed to the trained generator $G_T$, which would then generate the counterfactual next state $S^{'}_{t+1}$. These, together with $A^{'}_t$, would constitute the counterfactual database $\tilde{D}$, represented as
{\small
\begin{equation}
\begin{aligned}
\tilde{D} &= \{\tilde{D}_{0}, .., \tilde{D}_{i}, .., \tilde{D}_{N-1}\}=\{(s^{'}_{0}, a^{'}_{0}, .., s^{'}_{t}, a^{'}_{t}, .., s^{'}_{T-1})^{M-1}_{j=0}\}^{N-1}_{i=0},
\end{aligned}
\end{equation}
}
where $T$ represents the total number of steps, $M$ is the number of dialogues, and $N$ is the number of counterfactual databases. 


\subsubsection{Counterfactual Inference via Kernel Quantile Regression} \label{CI-KQR}
Traditional regression estimation focuses on estimating the conditional mean of $Y$ given $X$, typically represented by the function $f(X)$. On the other hand, quantile regression~\cite{Koenker2001} is concerned with estimating conditional quantiles, specifically the $\tau$-th quantile $\mu \tau$, which is the minimum value $\mu$ such that $P(Y\leq\mu|X)=\tau$, where $\tau$ is predefined. Quantile regression settings may encounter the quantile crossing problem~\cite{Takeuchi2009}, and a loss function is applied to learn \textit{all} conditional quantiles of a given target variable to address the problem~\cite{Tagasovska2019}. This is a non-trivial task. Fortunately, the framework in~\cite{Xie2024}, which we adapt here, only requires learning a single quantile to contain important information about the noise term. 

Interestingly, the work in~\cite{Xie2024} established a theoretical connection between counterfactual reasoning and nonlinear quantile regression that allows for bypassing the need for estimating both the SCM and noise values. 
Specifically, suppose $Y$ satisfies the SCM and $X, Z$, {\Large$\varepsilon_Y$} cause the outcome $Y$, and suppose further that the evidence observed in the data is $<X=x,Y=y,Z=z>$. We find that the counterfactual outcome of $Y$ if $X$ had been instead $x'$ given the observed evidence, denoted by $Y_{X=x'}|X=x,Y=y,Z=z$, is equal to the $\tau$-th quantile of the conditional distribution $P(Y|X=x', Z=z)$ where $Y=y$ is the $\tau$-th quantile of $P(Y|X=x, Z=z)$, i.e., $\tau=P(Y\le y|X=x, Z=z)$. Applying this theory to our work would now equate the counterfactual next state to
\begin{equation}
    S'_{t+1} = g(S_{t}, A'_{t}, \mathcal{\hat{L}}_{+1}).
    \label{eq:kqr}
\end{equation}
Hence, the counterfactual outcomes can be determined through quantile regression from purely factual observations or what can be recovered from them, making absent the noise estimates in Eq. (\ref{eq:kqr}) (cf: Eq. (\ref{eq:causal_func}) in which it is necessary to estimate the noise terms). 

\begin{figure}
\vspace{-3mm}
    \centering
    \includegraphics[width=0.95\linewidth]{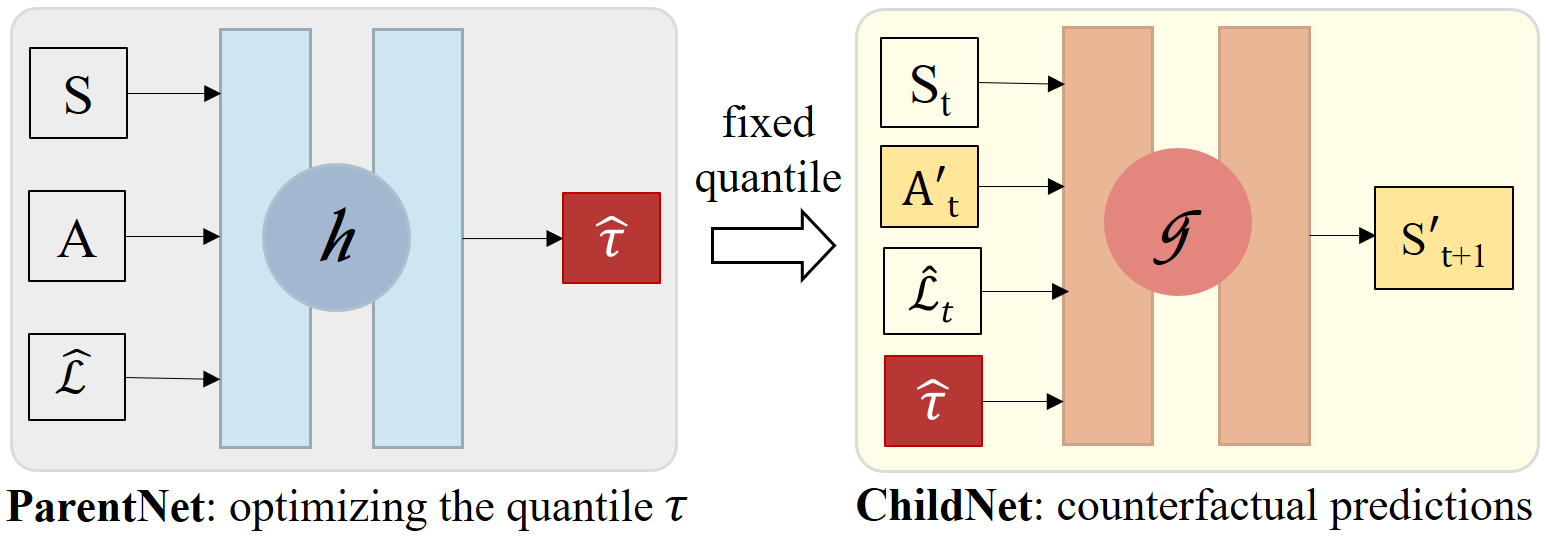}
\vspace{-3mm}
    \caption{KQR employs two networks: \textbf{ParentNet} optimizes $\tau$, yielding $\hat{\tau}$, which controls the  prediction of counterfactual next state $S'_{t+1}$ by \textbf{ChildNet} based on ($S_{t}$, $A'_{t}$, $\mathcal{\hat{L}}_{+1}$) as input.}
    \vspace{-5mm}
    \label{fig:kqr_model}
\end{figure}

The implementation of this method performs counterfactual inference using a bi-level optimization framework involving two networks: ChildNet for lower-level optimization and ParentNet for upper-level optimization, shown in Fig.~\ref{fig:kqr_model}. 
The ParentNet is a simple data-dependent model to learn the optimized quantile $\hat{\tau}=h(S, A, \hat{L})$, where $h$ is a neural network. Then, $\hat{\tau}$ is used as the input of the ChildNet on the lower-level problem to perform quantile regression by capturing the difference in different samples with a shared neural network. The ChildNet, on the other hand, is a model designed to perform counterfactual prediction of next states per Eq. (\ref{eq:kqr}), where $g(\cdot)$ is a conditional neural network that efficiently estimates quantiles for counterfactual inference across multiple samples. Like in BiCoGAN, $A'_{t}$ is produced based on the causal discovery results (in Section~\ref{retrieval_model}).
To sum up, this method learns the optimal quantile-based predictions for counterfactual analysis.

\subsection{Optimal Policy Learning} \label{policy}
Following the generation of the counterfactual dataset $\tilde{D}$, our framework then learns the policies on $\tilde{D}$ that would maximize future rewards. For this purpose, we choose the Dueling Double Deep Q-Network (D3QN)~\cite{raghu2017deep} (Fig.~\ref{fig:system_architecture}, in orange) to reduce overestimation bias in the Q-values
~\cite{mnih2013playing}. 
D3QN uses state-dependent advantage function $A(s, a)$ and value function $V(s)$: while $A(s,a)$ capitalizes on how much better one action is than the other actions, $V(s)$ indicates how much reward will be achieved from state $s$. The Q-value can be calculated over $\tilde{D}$ as follows:
{\small
\setlength{\abovedisplayskip}{5pt}  
\setlength{\belowdisplayskip}{5pt} 
\begin{equation}
    \begin{aligned}
    Q(s', a';\theta) = \mathbb{E}\left[r(s', a') + \gamma \max_{a'} Q(s', a';\bar{\theta}) \mid s^{'}, a^{'}\right],
    \end{aligned}
    \label{qlearning}
\end{equation}
}
\noindent where $r(s^{'}, a^{'})$ is the reward for taking counterfactual action $a^{'}$ at state $s^{'}$ in $\tilde{D}$, $\gamma$ is the discount factor of the maximum Q-value from the next state. $\bar{\theta}$ and $\theta$ are the weights of the target and main networks. The reward function $r(s^{'}, a^{'})$, which is calculated by an LSTM-based model trained on the dialogues and donation values, is used to compute the reward in ($s'_{t}$, $a'_{t}$) as follows:
{\small
\setlength{\abovedisplayskip}{5pt}  
\setlength{\belowdisplayskip}{5pt} 
\begin{equation}
  r(s^{'}, a^{'}) = 
  \begin{cases} 
    0, & \text{if } t < T-1, \\
    \text{DDP}(d) = \text{LSTM}(\text{BERT}(\{s'_{t}, a^{'}_{t}\}_{t=0}^{T-1})), & \text{otherwise}.
  \end{cases} 
  \label{reward_fun}
\end{equation}
}
\noindent where $T$ is the dialogue length in unit time, and $d$=$(s^{'}_{0}, a^{'}_{0}, s^{'}_{1}, a^{'}_{1}$, ..., $s^{'}_{t}, a^{'}_{t}$, ..., $s^{'}_{T-1})$ represents the sequence of states and actions. Target Q-values are derived from actions through a feed-forward pass on the main network, diverging from direct estimations from the target network. During policy learning, the state transition starts with state $s^{'}_{0}$, and the optimal counterfactual action $a^{*}_{0}$ is selected by computing $argmax_{a^{'}_{0i}}$ $Q(s^{'}_{0}, a^{'}_{0i}; \theta, \alpha, \beta)$, where $i$=0, 1, 2, .., $N-1$, and $N$ is the number of counterfactual databases. We use Mean Squared Error (MSE) as loss function to update the D3QN weights. 

The next step involves training policies on $\Tilde{D}$ to optimize the Q-values and improve the predicted future cumulative rewards over the ground truth. 
The reward (donation amount) for each dialogue is predicted from Eq. (\ref{reward_fun}) by performing a cumulative summation on the predictions of the trained reward model, as follows:
{\small
\begin{equation}
\begin{aligned}
  \hat{R}_{i} &= DDP(d) = LSTM(BERT(\{s'_{0}, a'_{0}, s'_{1}, a'_{1}, ..., s'_{T-1}\})), \\
  R_c &= \left[ \sum_{i=0}^{0} \hat{R}_i, \sum_{i=0}^{1} \hat{R}_i, \ldots, \sum_{i=0}^{k-1} \hat{R}_i, \ldots, \sum_{i=0}^{M-1} \hat{R}_i \right],
  \label{creward_fun}
\end{aligned}
\end{equation} 
}
where $M$ is the total number of dialogues, $R_c$ is the cumulative reward list over $M$ dialogues with each $R_c^k = \sum_{i=0}^{k-1} \hat{R}_i$ indicating the cumulative reward over $k$ dialogues.

\section{Experiment Setup} \label{experiment_setup}
We did not have to perform any field experiment, instead, we used the P4G dataset to validate our approach. 
P4G enables a structured analysis of persuasive strategies within realistic, dialogue-based contexts, enhancing research into effective persuasion techniques~\cite{Wang2019}.
This was populated from online tasks that consisted of surveys, persuasion dialogues, and donation confirmations. The 1,285 individuals who performed the tasks had to answer surveys, among which are their responses to a personality inventory (25 questions)~\cite{Goldberg1992} that offered us the opportunity to estimate the OCEAN values. Participants were randomly assigned to be persuader or persuadee: while persuaders were provided with tips on different persuasion strategies, persuadees only knew they would talk about a specific charity. Participants were encouraged to continue the conversation until an agreement to donate or not was reached. Refer to the work in~\cite{Wang2019} for further details on P4G.

We implemented our models using PyTorch and trained them on a GeForce RTX 3080 GPU (10G). The model configurations used in the experiments are as follows: (1) TP3M: 1024 hidden units, a batch size of 64, learning rate of 0.0001, and 100 epochs. (2) BiCoGAN: 100 hidden units, a batch size of 100, learning rate of 0.0001, and 10 epochs. (3) RB: 256 hidden units, a batch size of 64, learning rate of 0.0001, and 1,000 epochs.
D3QN: 256 hidden units, a batch size of 60, learning rate of 0.001, and 20 epochs. (4) KQR: A batch size of 200 with a learning rate of 0.1 for ParentNet and 0.001 for ChildNet.
For BiCoGAN, we set the dimension of the noise term to be the same as BERT's embedding of 768. 
To reduce potential bias from outliers and ensure a more balanced training of the models, we set any donation exceeding \$10 to be \$10. 

We generated 50 counterfactual dialogues using different sets of counterfactual actions. 
Each dialogue showcases flexible-length exchanges between EE and ER, with alternating utterances. 
We did an 80/20-split of the data for training/testing. All models were optimized using Adam~\cite{kingma2014adam} with a 0.0001 learning rate. We also set D3QN's discount factor $\gamma$ to 0.9. For the KQR, we use a decay factor (controls the smoothness of the kernel) of 0.9 for the ParentNet and apply a QP solver~\cite{cvxopt} during training for the ChildNet. 



\section{Discussion of Results and Analyses} \label{results}

\subsubsection{Predictions of the latent personality traits.}
\begin{figure}
\vspace{-3mm}
    \centering
    \includegraphics[width=0.95\linewidth]{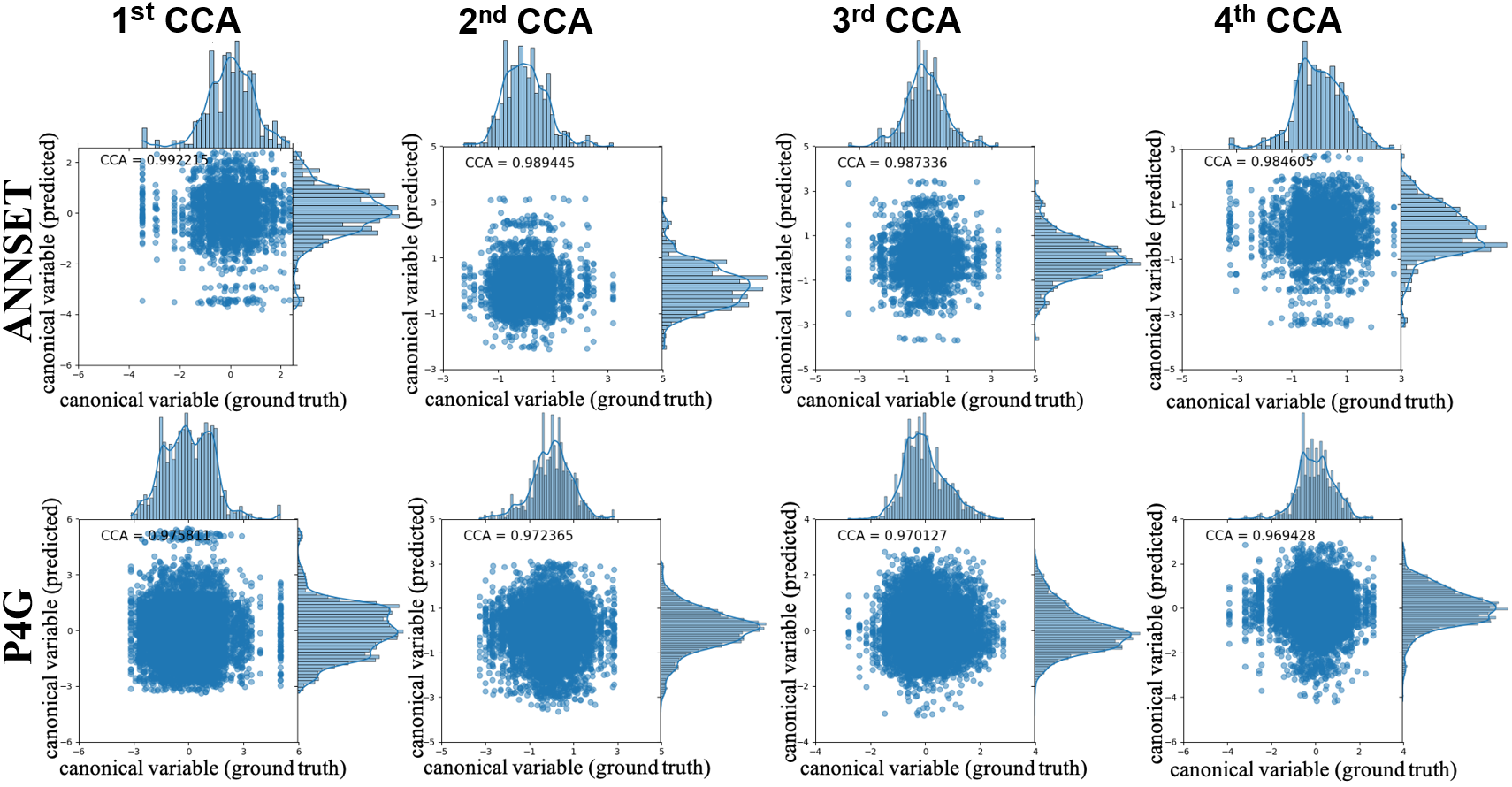}
    \vspace{-3mm}
    \caption{Top four CCA correlations, along with their distributions, verify the strong correlation between the ground truth and the TP3M-predicted OCEAN values.}
    \vspace{-6mm}
    \label{fig:cca_component}
\end{figure}

TP3M takes as input a one-turn utterance composed of state-action pair ($s_{t-1}$, $a_{t-1}$), to predict the persuadee's personality at time $t$.
We use five-fold cross-validation to assess TP3M's reliability to ensure its generalizability to unseen data.
The results are as follows: MSE=0.166, RMSE=0.407, MAPE=0.092, R$^2$=0.830, and MAE=0.254~\footnote{MSE - Mean Squared Error; RMSE - Root Mean Squared Error; MAPE - Mean Absolute Percentage Error; R$^2$ - Coefficient of Determination; MAE - Mean Absolute Error},
which indicate that TP3M performs well in minimizing prediction errors. To further assess TP3M's performance, we computed the top four canonical correlation analysis (CCA) components of the ground-truth OCEAN and TP3M’s estimated OCEAN values (Fig.~\ref{fig:cca_component}). 
Interestingly, these top four CCA correlations exceed 0.96, which indicates that analyzing the ongoing dialogue can reveal latent personality factors closely tied to the persuadee's personality.


\subsubsection{Causal relations among utterance strategies} \label{causal_rel}
\begin{figure}
\vspace{-3mm}
    \centering
    \includegraphics[width=0.95\linewidth]
    {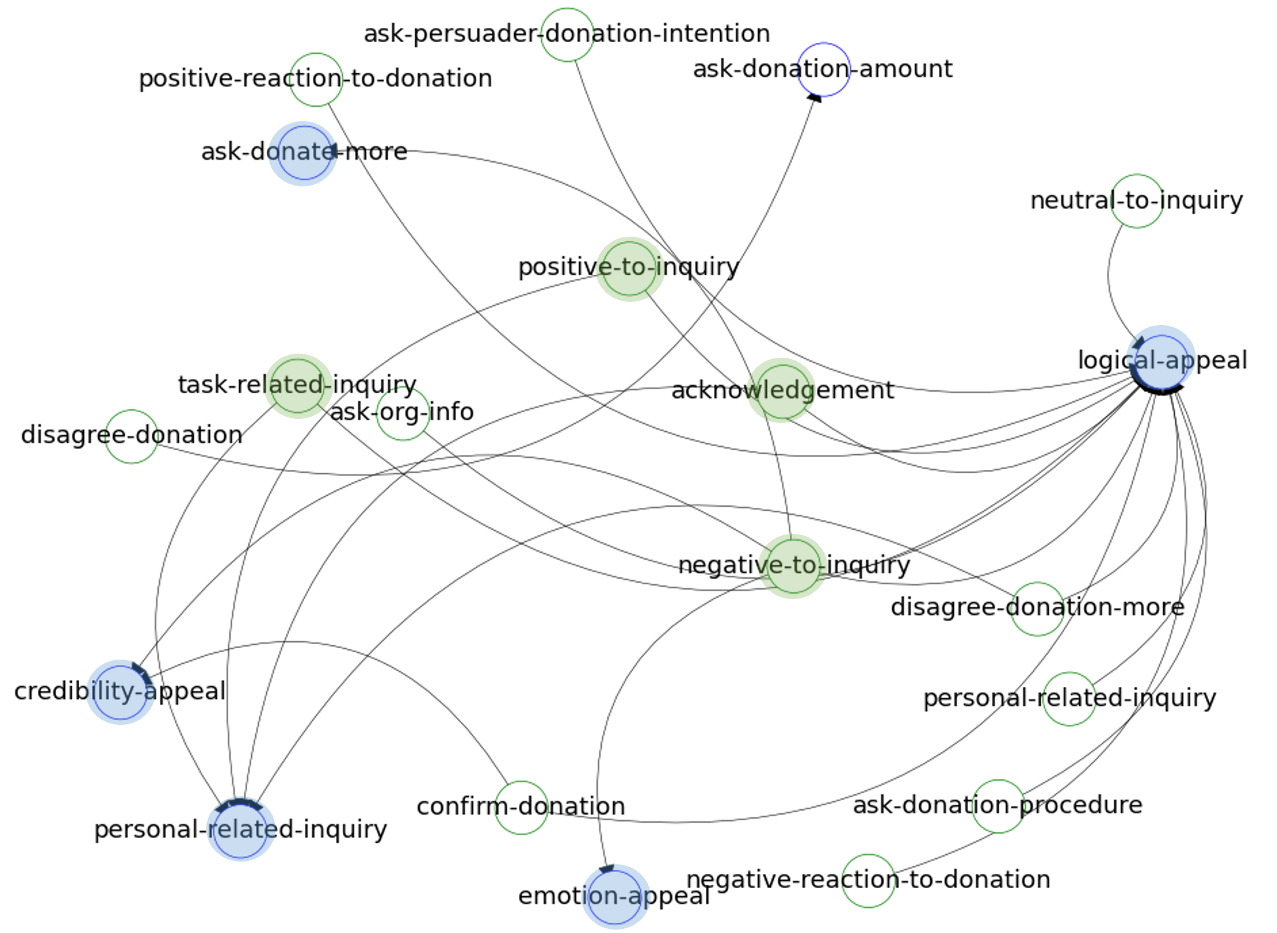}
    \vspace{-3mm}
    \caption{Discovered causal relations among strategies using GRaSP. While EE strategies (green) are causes, ER strategies (blue) represent the outcomes of EE's influence over ER. 
    } 
    \vspace{-6mm}
    \label{fig:causal_graph}
\end{figure}

GRaSP discovered 36 cause-effect pairs that are plausibly behind the utterance strategies of EE and ER, where EE strategies serve as causes and ER strategies as effects.
A portion of the causal graph discovered by GRaSP is shown in Fig.~\ref{fig:causal_graph},
which demonstrates interestingly how various strategies used by EE could influence the strategies of ER.
Following are some key insights.  
(1) Prevalence of \textit{logical appeal} (18 cause-effect pairs): this strategy was predominantly used by ER in response to EE's \textit{positive-to-inquiry} and \textit{task-related-inquiry} strategies. Our statistics show that 238 dialogues feature \textit{logical-appeal} and \textit{positive-to-inquiry} strategies together, and 271 dialogues feature \textit{logical-appeal} and \textit{task-related-inquiry} strategies. These highlight \textit{logical appeal}'s key role in addressing EE's informational needs.
(2) Role of \textit{personal-related-inquiry} (4 pairs): this strategy is frequently used by ER following EE's \textit{positive-to-inquiry} and \textit{acknowledgment}. We find that 263 dialogues feature \textit{personal-related-inquiry} and \textit{positive-to-inquiry} strategies together, and 172 dialogues feature \textit{personal-related-inquiry} and \textit{acknowledgment} strategies, highlighting the role of \textit{personal-related-inquiry} in personalizing the dialogues. (3) Role of \textit{emotion-appeal} (7 pairs): ER frequently employs this strategy when responding to EE’s \textit{negative-to-inquiry} strategies, with 265 dialogues demonstrating the interplay between \textit{emotion-appeal}
and \textit{credibility-appeal}, to foster an emotional connection particularly when logical arguments are inadequate.

\subsubsection{Increased cumulative persuasion outcomes (donation amounts)} \label{cumrew_preds} 
We use our models to compute the cumulative predicted rewards, i.e., donation amounts, for two ground truth cases and eight counterfactual inference variants. The two ground truth cases are (1) {\texttt{GT}} - actual donation amounts 
in the P4G dialogues, and (2) {\texttt{Pred\_GT}} - donation amounts 
predicted by the DDP model from P4G. The eight variants come from BiCoGAN and KQR, as explained below. 

Recall that to facilitate the creation of the counterfactual database $\Tilde{D}$, we introduce the counterfactual action set~\{$a^{'}_{t}$\} as input to the counterfactual inference methods. We select the counterfactual actions using two different approaches: (1) randomly choose from ER's utterances or (2) guide the selection using the discovered causal relations. 
The initial step in (1) involves creating \{$a^{'}_{t}$\} by randomly selecting in P4G the actual action set \{$a_{t}$\}. To prevent greeting utterances from appearing in the middle or at the end of the dialogues when learning the policy for selecting \{$a^{'}_{t}$\}, we exclude utterances with \textit{greeting} strategy when randomly selecting utterances for \{$a^{'}_{t}$\}. On the other hand, for (2), GRaSP discovered the cause-effect pairs.
The variables {\texttt{BiCoGAN\_random}},
{\texttt{BiCoGAN\_CD}},
{\texttt{BiCoGAN(latent)\_random}}, and
{\texttt{BiCoGAN(latent)\_CD}} indicate the
cumulative rewards coming from BiCoGAN variants (\texttt{CD} refers to causal discovery via GRaSP, and \texttt{latent} to TP3M's estimates of OCEAN); and
(2) {\texttt{KQR\_random}}, {\texttt{KQR\_CD}}, {\texttt{KQR(latent)\_random}}, and {\texttt{KQR(latent)\_CD}} are the
cumulative rewards from KQR variants.

\begin{figure}[t]
\vspace{-3mm}
    \centering
    \includegraphics[width=0.8\linewidth]{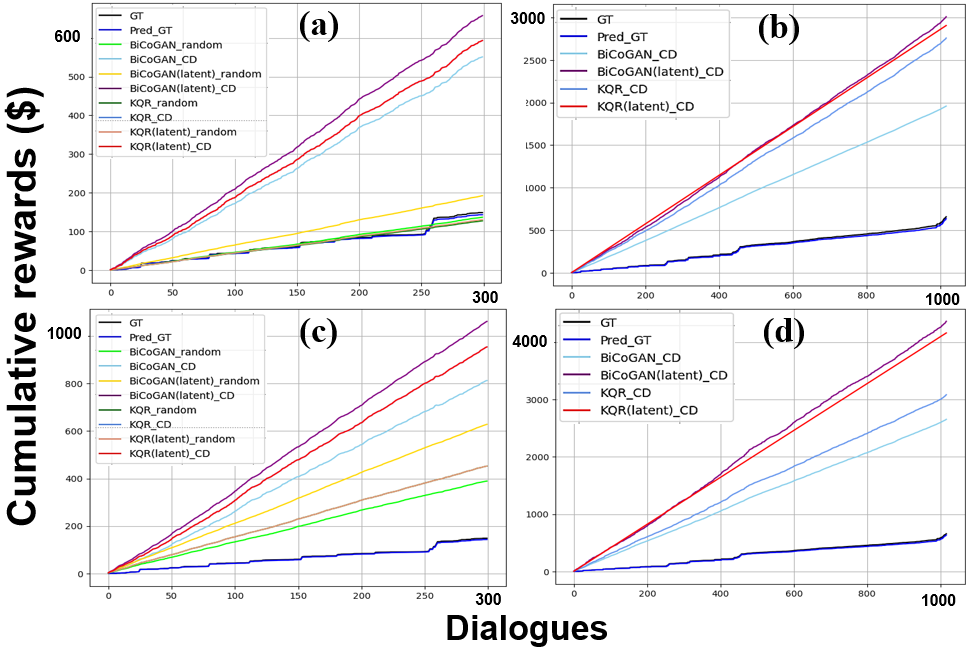}
    \vspace{-2mm}
    \caption{
    Cumulative rewards $DDP(\tilde{D})$ for (a, b) and $DDP(D*)$ for (c, d); using ANNSET in (a, c) and the full P4G in (b, d).
    }
    \vspace{-5mm}
    \label{fig:coarse_grained_300_full}
\end{figure}

In Figs.~\ref{fig:coarse_grained_300_full}(a) and ~\ref{fig:coarse_grained_300_full}(b), we show the cumulative rewards for \texttt{GT}, \texttt{Pred\_GT}, 
and the eight counterfactual inference variants.
The cumulative reward predictions of the eight variants
for $N=50$ counterfactual databases
are computed as {\small $\overline{DDP(\tilde{D})} = \frac{1}{N} \sum^{N-1}_{i=0} DDP(\tilde{D}_{i})$}. 

First, we can observe that the cumulative rewards from {\texttt{GT}} and {\texttt{Pred\_GT}}
are nearly overlapping and the total donations are very close. This means that the DDP model is highly accurate in its predictions of the ground truth cumulative rewards. This is very important as it gives credence to the next set of cumulative rewards that would come from the eight variants, in which there is significant improvement.
Observing deeper, Fig. 7(a) shows the following results can be observed for ANNSET:
(1) There is a significant increase from {\texttt{BiCoGAN\_random}} (\$487.09) to {\texttt{BiCoGAN\_CD}} (\$573.68), which clearly indicates the contribution of the causal relations. (2) The addition of OCEAN estimates further increased the cumulative rewards, {\texttt{BiCoGAN(latent)\_random}} (\$700.97) and {\texttt{BiCoGAN(latent)\_CD}} (\$753.20), contrast to when OCEAN is not estimated. This supports our hypothesis that there is value in accounting for latent factors in the state transition dynamics (Fig.~\ref{fig:theoretical_schema}).

Focusing next on the cumulative rewards from the KQR variants in the same way, we can observe that (1) while there is also a significant increase in cumulative rewards from {\texttt{KQR\_random}} (\$562.27) to {\texttt{KQR\_CD}} (\$697.55), (2) there is almost no perceivable increase when the latent personality estimates were included: {\texttt{KQR(latent)\_ random}} (\$562.35) and {\texttt{KQR(latent)\_CD}} (\$697.58). 
This tells us that for KQR, while leveraging causal knowledge is also significantly beneficial for counterfactual inference, the latent factors were unnecessary. The addition of OCEAN estimates contributed very little to improve the persuasion outcomes, and noise terms need not be estimated. Overall, however, the benefit is greater when these latent factors are estimated, as demonstrated by the BiCoGAN variants.   

We then observe in Fig.~\ref{fig:coarse_grained_300_full}(b) the results of using the full P4G but without the \texttt{random} variants. 
We know that the results are better when causal discovery is employed, as opposed to randomly selecting the counterfactual actions. Furthermore, in real-world scenarios, randomly choosing an utterance for ER as response to EE can be risky and lead to ineffective communication. 
Hence, we exclude the \texttt{random} variants to represent more realistic behavior by the system. Fig. 7(b) shows significant increase with causal discovery, {\texttt{BiCoGAN\_CD}} (\$1958.12) and {\texttt{KQR\_CD}} (\$2757.70), let alone when the latent OCEAN estimates were integrated, {\texttt{BiCoGAN(latent)\_CD}} (\$3008.50) and {\texttt{KQR(latent)\_CD}} (\$2905.03). This relative increase in cumulative rewards, compared to those in Fig.~\ref{fig:coarse_grained_300_full}(a), is because as the dataset becomes larger, the causal discovery learns more effectively the causal relations from more personalized interactions.


We see in Figs.~\ref{fig:coarse_grained_300_full}(c) and~\ref{fig:coarse_grained_300_full}(d) the cumulative rewards $DDP(D*)$, $D*$ being the dialogues derived using the optimal policy learned by D3QN from the counterfactual dataset $\tilde{D}$. We see again the accurate prediction of the ground truth cumulative rewards, which once again supports the idea that the other cumulative reward predictions can also be considered reliable. Interestingly, the cumulative rewards behave in the same fashion:
(1) allowing the discovered causalities to guide the counterfactual inference,  
{\texttt{BiCoGAN\_CD}} and {\texttt{KQR\_CD}}, significantly improves the persuasion outcomes, and (2) there is further significant increase in persuasion outcome when OCEAN is estimated, as in {\texttt{BiCoGAN(latent)\_random}} and {\texttt{BiCoGAN(latent)\_CD}}. (3) But the cumulative rewards from KQR with and without OCEAN estimates are practically overlapping. 

Overall, therefore, our results show that identifying strategy-level causal relations from observed data can generate counterfactual system actions that improve persuasive dialogue outcomes. Additionally, accounting for latent factors - personality and noises - further enhances the creation of counterfactual data, offering strategic counterfactual utterances that could lead to increased rewards. 

\subsubsection{Quality of the generated counterfactual data}
\begin{figure} [t!]
\vspace{-3mm}
    \centering
    \includegraphics[width=\linewidth]{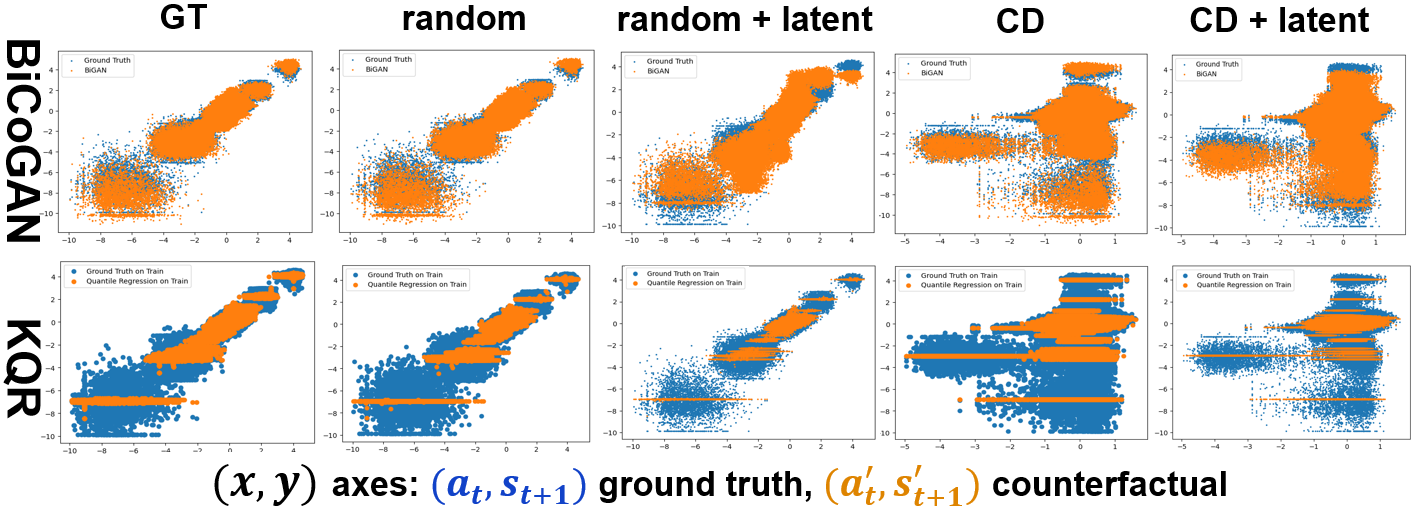}
    \vspace{-6mm}
    \caption{Observed ground truth in ANNSET and the generated counterfactual actions and next states.
    } 
    \vspace{-6mm}
    \label{fig:cf_relations}
\end{figure}

Our results show
that BiCoGAN outperforms KQR (Fig.~\ref{fig:coarse_grained_300_full}). With a sharper lens, we look at the generated $A'_t$ and $S'_{t+1}$, and their relation that emerged from using the BiCoGAN and KQR variants. We see in Fig.~\ref{fig:cf_relations}(top) that with BiCoGAN, in contrast to KQR, a significant number of $S'_{t+1}$ data points (yellow) closely align with the ground truth $S_{t+1}$ (blue). This suggests that the generated alternative reality can still be identifiable (does not unrealistically diverge) from real-world conditions. We thought initially that KQR efficiently accounts for latent noises even without estimating them, hence, the counterfactual data it generates is sharply and effectively focused only on pertinent aspects of the ground truth, seen as limited yellow coverage in Fig.~\ref{fig:cf_relations}(bottom). However, this did not square with the persuasion outcomes resulting from the KQR variants (Fig.~\ref{fig:coarse_grained_300_full}).

\section{Shaping Ethics in Generative AI}
Our work underscores the ethical and practical implications of generative AI in persuasive technology (GenAI-PT). This study demonstrates that when applied responsibly to socially beneficial purposes, GenAI-PT can be a force to drive positive change (e.g., increased charitable donations). 
The potential for harm, such as manipulative persuasions, is mitigated by focusing on social good. Furthermore, trust can be achieved through counterfactual explanations (see~\cite{Del2023}), i.e., people can familiarize themselves with unknown processes by understanding the hypothetical conditions under which the outcome changes.
Our method is also unbiased as it does not overly rely on personality stereotypes, some of which may not be true, thus, can be harmful. Instead, it dynamically fine-tunes its personality trait estimates based on the ensuing utterances. 

While theory-centric approaches often risk homogenization and overlook diverse user contexts, our framework ensures adaptability to personal and contextual scenarios. By accounting for latent influences and users' shifting needs, we prioritize towards long-term relevance and user well-being. Future enhancements of our work should enable continual learning and self-improvement to foster enduring, trust-based relationships~\cite{Del2023} with users (e.g., in~\cite{Legaspi2021}).

\section{Addressing Critical Aspects of the Reviews}

Our work models behaviors in persuasive dialogues, i.e., how a persuasive system can dynamically adapt its responses to the evolving states of its users.
It aligns with data-driven, computationally-intensive theory development, and began with a hypothesis that is grounded on strong theoretical foundations of causal discovery, counterfactual inference, and latent knowledge estimation~\cite{Pearl2009,Spirtes2016,Bareinboim2022}.
However, instead of relying on mathematical proofs or theorems that may obscure the generative mechanism, we tested our hypothesis using real-world dialogue data.
While we evaluated on a (static) dataset, we can argue, that our framework enables the simulation of dynamic scenarios by estimating outcomes under alternative conditions that were not explicitly observed in the dataset. Furthermore, by training on the recorded dialogues and learning reward-based decision-making, the deep reinforcement learner can simulate a dynamic setting where actions influence future states even when the dataset itself is fixed. This enables generating trajectories that mimic real-world dynamism. Thus, even when real interactions are not possible, our framework can generate plausible sequences of events, which is practically useful especially when experimentation in the field becomes prohibitive.
As a result, beyond validating our hypothesis, we uncovered novel insights, such as leveraging causal and latent knowledge mitigates concerns about counterfactuals being out-of-distribution, and explicitly estimating latent factors proves more optimal even when avoidance is possible.

Our pipeline, albeit massive, consists of methods pieced together naturally but not without much deliberation or background information. Due to space constraints, we refer to the \href{https://github.com/ZenzenDatabase/PersonPersuasion}{\textbf{Supplementary Materials}} in which we articulated the considerations guiding our method selection. What was imperative was to demonstrate the added effect of the hypothesis-related components (causal, counterfactual and latent factor knowledge) when compared to another generative framework~\cite{zeng2024counterfactual} that we treated as the control/baseline version. In effect, the baseline provided for us a simple, straightforward and effective basis for comparison to prove our hypothesis and demonstrate its benefits. 

Indeed, while our results demonstrate optimized persuasion outcomes, more can be done to achieve full optimization. Given that the components are modularized, future work may explore alternative methods to enable further optimization. Additionally, optimizing across all possible counterfactual cases remains an open challenge for future work. For one, we could not explore all cases due to the computational complexity of our pipeline. Another is that randomness in certain instances (e.g., see  Section~\ref{cumrew_preds}) remains unavoidable in our methodology, which makes exhaustive enumeration of every scenario difficult. 

On the psychological aspect, we recognize the argument that the Big Five are inherently broad and applicable across diverse contexts, i.e. ``generic''; and that a theory-driven approach, rather than a purely data-driven one, might have identified psychological constructs more specific to persuasion as established in related literature. Indeed, and in fact, the work in~\cite{Wang2019} that constructed the P4G and used it, observed that more agreeable participants are more likely to donate — though not without a caveat. The authors warn that due to the relatively small dataset, the results are not statistically significant and caution against over-interpreting these effects until more annotated data become available. Thus, we deemed such inconclusive result to be insufficient for providing a theory-driven basis for structured exploration. 
Rather than applying theory-imposed static constructs, our model allows for emergent personality inference from natural language behavior, which aligns with contemporary views on the fluidity of personality expression (noteworthy surveys in~\cite{Kuper2021,Jayawickreme2021}), for instance, personality as a dynamic system emphasizing individual difference parameters describing fluctuations in personality \textit{states} contingent on contextual factors~\cite{Fleeson2001,Andersen2016,Beckman2017,Sosnowska2019}. We therefore posit that our adaptive process via TP3M is both psychologically motivated and empirically grounded. It is also important to emphasize that the Big Five have been described as aggregates of behavioral tendencies within finite multidimensional spaces~\cite{Goldberg1992, Costa2017}. Aggregation into broad factors offers key benefits, such as simplifying personality descriptions (parsimony) and enhancing the consistency and stability of outcome predictions~\cite{Baumert2017}.
It allows researchers to examine broad personality dimensions and their correlations with various aspects of human behavior. For instance, it has been found that personality factors, albeit aggregated scores, affect belief change, with conscientious, open and agreeable people being more convinced by emotional persuasion~\cite{Lukin2017}.


We also understand the concern about the very technical nature of our paper. We used detailed formulations to ensure the reproducibility of our work, allowing others to follow the methodology without having to additionally refer to other papers.
With an X-ray-like focus, Fig.~\ref{fig:system_architecture} reveals the internal composition of each component in our pipeline and the flow of data between them, thereby offering a level of detail beyond that of a typical flowchart.
Further on the issue of reproducibility, we also included in the \href{https://github.com/ZenzenDatabase/PersonPersuasion}{\textbf{Supplementary Materials}} details (e.g., data extraction and preprocessing) to help reproduce our generative framework. Although we also made transparent a great amount of detail (e.g., formulations, algorithms, hyperparameter tuning, model training, etc.) in Section~\ref{experiment_setup}, we also make our codes available in the~\href{https://github.com/ZenzenDatabase/PersonPersuasion}{\textbf{Supplementary Materials}}.

Regarding real-world deployment and scalability, we mentioned in the paper that it is not within the scope (for now) to realize a deployable system and interface, only the generative intelligence that would enable the system to be more personal and adaptive. When a system scales, its performance can change due to new conditions, unforeseen interactions, or evolving user behavior. Our causality-based counterfactual reasoning can provide insights into how the system might adapt to these changes by simulating different scenarios of scaling. This can guide the design of adaptive mechanisms that optimize system performance at scale.

Finally, we believe our work can be applied broadly in new areas. A fundamental problem is that many existing methods rely heavily on techniques that prioritize statistical patterns, disregarding the true causal mechanisms underlying complex real-world system dynamics. This can result to unreliable conclusions influenced by spurious correlations or random chance. We can see the relevance of overcoming this fundamental issue in emerging and exciting state-of-the-art~\cite{Mitra2022}.
Furthermore, tailoring our framework to support tasks such as recommendation, health intervention, and negotiation presents an exciting avenue for future work.

\bibliographystyle{ACM-Reference-Format}

\bibliography{references}



\end{document}